\documentclass[galaxies,article,accept,moreauthors]{mdpi}

\usepackage{upgreek}
\firstpage{1} 
\pubvolume{1}
\issuenum{1}
\articlenumber{0}
\pubyear{2026}
\copyrightyear{2026}
\externaleditor{Firstname Lastname} 
\datereceived{}
\daterevised{}
\dateaccepted{}
\datepublished{ } 

\usepackage{gensymb}
\usepackage{graphicx}

\newcommand{\ha}{H$\alpha$}
\newcommand{\oiii}{[O\,{\sc iii}]}

\newcommand{\nii}{[N\,{\sc ii}]}

\newcommand{\arcmin}{$^{\prime}$}
\newcommand{\arcsec}{$^{\prime\prime}$}
\newcommand{\kms}{\,km\,s$^{-1}$}	
\newcommand{\ngc}{NGC~6563}	
\newcommand{\degr}{$^\circ$}
\newcommand{\farcs}{\,.\!\!^{\prime\prime}}
\newcommand{\fs}{\,.\!\!^{\mathrm{s}}}

\Title{Morphokinematic structure of the Planetary Nebula NGC 6563}

\Author{Zahra Al$^1$\orcidN{}, 
Federico Soto-Badilla$^2$\orcidK{}, 
Y\"uksel Karata\c{s}$^3$\orcidM{},
Gerardo Ramos-Larios$^4$\orcidG{}, and
Roberto Vázquez$^{2}$\orcidA{}$^{*}$}

\AuthorNames{Zahra Al, Federico Soto-Badilla, Y\"uksel Karata\c{s}, 
Gerardo Ramos-Larios, Roberto Vázquez}

\address{%

$^{1}$ \quad Institute of Graduate Studies in Science, Programme of Astronomy and Space Sciences,  {\.I}stanbul University, 34116, \"Universite-Istanbul, T\"urkiye; zahraa96al@gmail.com (Z.A.)\\
$^{2}$ \quad Instituto de Astronomía, Universidad Nacional Autónoma de México, 22860 Ensenada, B. C., Mexico; fsoto@astro.unam.mx (F.S.-B.), vazquez@astro.unam.mx (R.V.)\\
$^{3}$ \quad Department of Astronomy and Space Sciences, Science Faculty, {\.I}stanbul University, 34116, \"Universite-Istanbul, T\"urkiye; karatas@istanbul.edu.tr (Y.K.)\\
$^{4}$ \quad Instituto de Astronom\'{i}a y Meteorolog\'{i}a, CUCEI, Universidad de Guadalajara,  44130 Guadalajara, Jal., Mexico; gerardo@astro.iam.udg.mx (G.R.-L.)}

\corres{Corresponding author: vazquez@astro.unam.mx (R.V.)}

\abstract{We present a morphokinematic analysis based on high-resolution long-slit echelle spectroscopy of the \nii$\lambda6583$ line and narrowband imaging. Position–velocity diagrams reveal asymmetric expansion and localized kinematic features. We derive a systemic velocity of $V_{\rm sys}^{\rm LSR} = -25\pm1$\kms\ ($V_{\rm sys}^{\rm HEL} = -34 \pm 1$\kms) and a main shell expansion velocity of $V_{\rm exp} = 22 \pm 1$\kms. Three-dimensional modeling indicates an ellipsoidal main body surrounded by a thin shell, two ear-like protrusions, and additional small-scale structures. The corresponding kinematic ages are $3\,600 \pm 700$ yr for the ellipsoid and ring, and $7\,500 \pm 1\,000$ yr and $8\,800 \pm 1\,500$ yr for the two opposite ear-like protrusions, respectively, indicating that these outer structures predate the main nebular envelope. The kinematic asymmetry and enhanced emission regions suggest evolution within a non-uniform ambient medium. At the same time, the presence of collimated ear-like structures is consistent with shaping influenced by binary interaction, where earlier outflows preceded the ejection of the dense shell. NGC\,6563 therefore appears to be a dynamically evolved system shaped by the combined effects of episodic mass ejection and environmental interaction.}

\begin{document}



\section{Introduction}
\label{intro}

Planetary nebulae (PNe) represent a crucial stage in the late evolution of intermediate-mass stars. The study of PNe morphology has gained significant attention in astrophysical research, as it extends beyond simple classification to serve as a key tool for understanding their origin, structure, and evolution \citep{Curtis18}. Over time, various morphological classification schemes have been proposed \citep[e.g.,][]{Khromov68, balick87, Aaquist96, Manchado1996, Parker2006}, classifying PNe into categories such as spherical, elliptical, bipolar, quadrupolar, and more complex or irregular forms (e.g., butterfly-shaped nebulae). Several mechanisms have been suggested to explain these morphologies, including the rotation of the progenitor star, the influence of magnetic fields, and the presence of binary star systems \citep{kwok24}. Moreover, a comprehensive understanding of planetary nebula (PN) morphologies requires consideration of the influence of the interstellar medium (ISM) on their kinematics and dynamics, as well as the role of jets and outflows \citep[e.g.,][]{ism, knots, muller, sabin, akashi, vazquez, markito, 2025MNRAS.541.3932F}.

\begin{figure}[H]
\includegraphics[width=\textwidth]{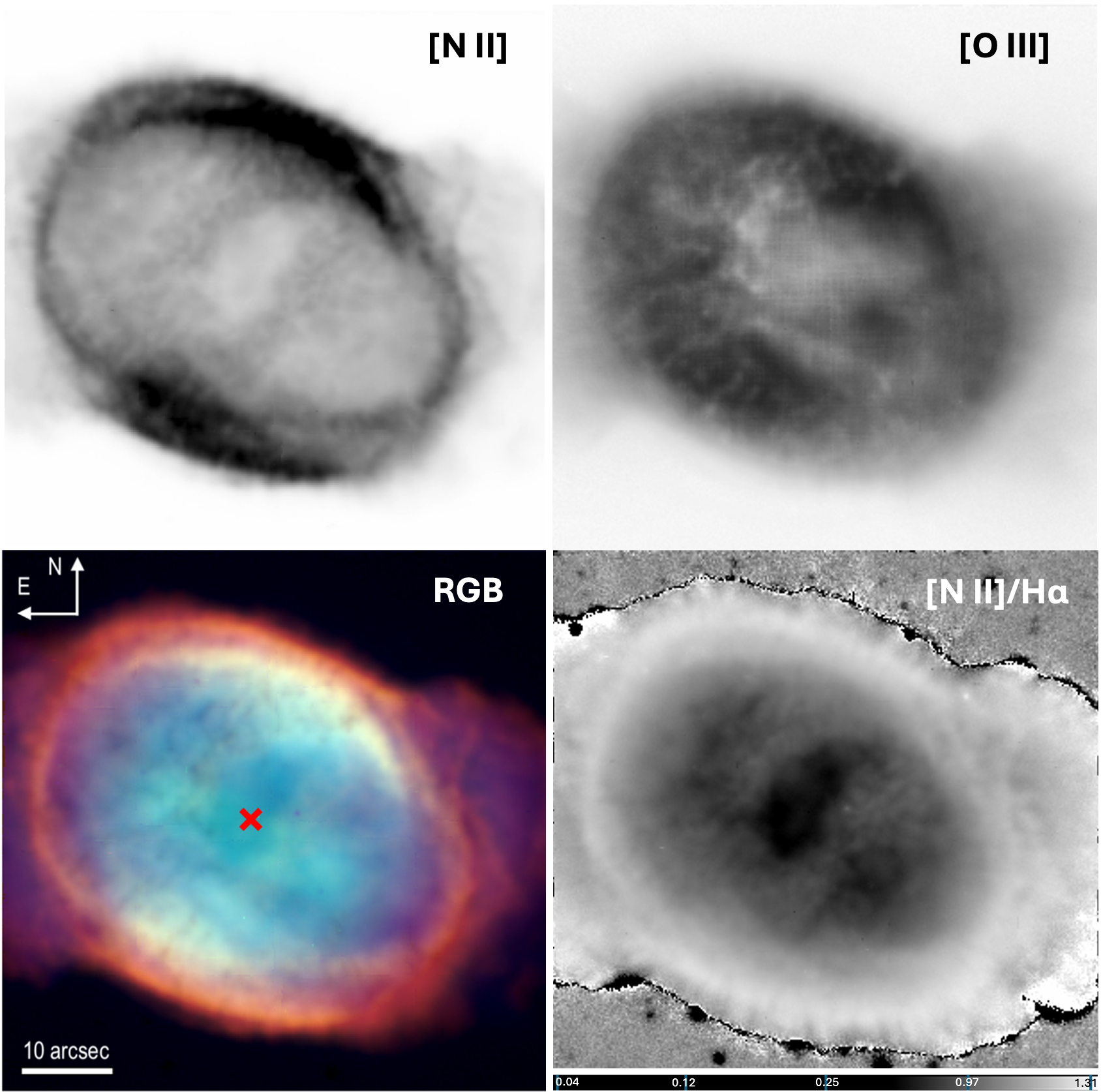}
\caption{Optical images of {\ngc}. 
{\it Top:} MUSE continuum-subtracted images in {\nii} (left) and {\oiii} 
(right), shown in grey scale. 
{\it Bottom left:} RGB composite image showing the spatial distribution of 
the main nebular emission lines, with {\nii}, {\ha} and {\oiii} assigned to 
the red, green and blue channels, respectively. The red cross marks the 
Gaia DR3 position of the central star. 
{\it Bottom right:} {\nii}/{\ha} ratio map. 
The images were obtained 
with MUSE assisted by the Adaptive Optics Facility on ESO's Very Large 
Telescope and were processed by the authors from publicly available archival 
data described in Section~2.
}
\label{rgb}
\end{figure} 

Located in the constellation Sagittarius, with its equatorial coordinates of $\alpha(2000)=18^{\rm h}12^{\rm m}2\fs59$ and $\delta(2000)=-33$\degr52\arcmin$6\farcs6$, {\ngc} (PN G358.5-07.3) was discovered by James Dunlop in 1826 using his 23-in aperture and 274-cm long telescope \citep{Dunlop}. 

NGC\,6563 has an elliptical morphology, with one side narrower than the other, giving it an egg-like appearance, as can be seen in Figure~\ref{rgb} \citep[see][]{Curtis18, Khromov68, Greig, Stan2010, Bouvis2025}. This nebula also exhibits two small opposite lobes (`ears') \citep{schwarz92, Akashi2021}. These ears differ from bipolar lobes in that bipolar lobes widen outward, whereas ears become narrower (Figure~\ref{OmegaCAM}). In addition, the ears are smaller than the main nebular body, while bipolar lobes typically extend beyond the main nebula \citep{Akashi2021}. 

In this study, we present the first morpho-kinematic analysis of {\ngc}. Since projection effects can significantly bias the interpretation of nebular structure, three-dimensional morphokinematic modelling is required to recover the intrinsic geometry and velocity field of PNe \citep{steffen}. We analyze its spectra and construct a 3D model consistent with its observed spectral properties. This allows us to identify the main structural components, estimate their kinematic ages, and explore the physical mechanisms shaping the nebula.

\begin{table}
\centering
\caption{Astrometric parameters (parallax and proper motion components) and $G-$mag for Central Star of \ngc. Source ID 4039600536544395392.}
\renewcommand{\arraystretch}{1.5}
\begin{tabular}{lc} 
\toprule
 $\varpi$ (mas)   & $1.07 \pm 0.13$ \\ 
 $d_{\varpi}$ (parsec)  & $934 \pm 112$  \\ 
 $\mu_{\alpha}$ (mas yr$^{-1}$) & $1.11 \pm 0.15$  \\ 
 $\mu_{\delta}$ (mas yr$^{-1}$) & $-3.10 \pm 0.11$  \\ 
 $G$ (mag)  & $17.276 \pm 0.002$ \\ 
\toprule
\end{tabular}
\label{tab:params}
\end{table}

\section{Observations}
\label{sec:errors}
\noindent 

\subsection{Imagery}

In order to obtain a detailed view of the internal structure of NGC~6563 
(Figure~1), we used integral field spectroscopy (IFS) data obtained with 
the Multi Unit Spectroscopic Explorer (MUSE), mounted on Unit Telescope 4 
(UT4, Yepun) of the Very Large Telescope (VLT). The observations were 
carried out on 2017 July 16 as part of programme ID 60.A-9100(H), led by 
the MUSE team, in Wide Field Mode with adaptive optics (AO) correction. 
The total exposure time was 480\,s, covering a field of view of 
1\arcmin$\times$1\arcmin\ and a spectral range from 4700 to 9350\,\AA. 
This setup provides a spatial sampling of 0.2\arcsec\ per pixel.

From the MUSE datacube, we extracted continuum-subtracted images in the 
main emission lines \ha, \nii, and \oiii. These images were used 
both individually and in combination to inspect the ionization-dependent 
morphology of the nebula. In particular, the \nii\ image traces the 
low-ionization structures analysed in this work, whereas the \oiii\ 
image highlights the higher-excitation inner nebular emission. We also 
constructed an RGB composite image, combining \nii, \ha, and 
\oiii, to provide a global view of the spatial distribution of the 
different ionization zones. Finally, we produced a {\nii}/{\ha} ratio 
map, which helps to minimize purely density-related features and to 
emphasize spatial variations associated with excitation and/or abundance. 
This ratio map is useful for comparing the low-ionization structures with 
the rest of the nebular shell and for assessing whether localized 
enhancements may be related to excitation effects rather than only to 
surface-brightness variations.

\begin{figure}[H]
\includegraphics[width=0.9\textwidth]{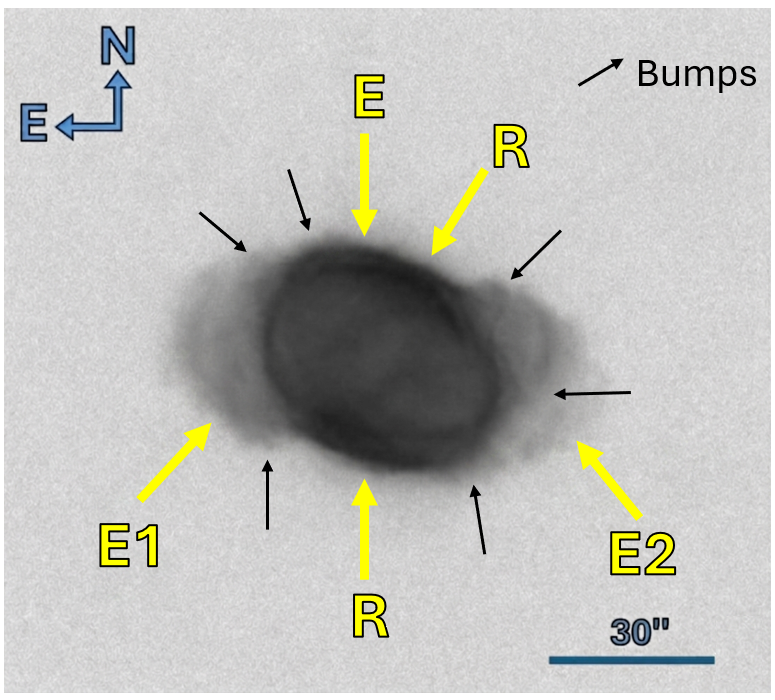}
\caption{Archival VST/OmegaCAM narrow-band image of NGC 6563 obtained with the NB-659 filter which samples the {\ha}+{\nii} emission region. Stellar PSFs have been subtracted to enhance the visibility of faint nebular structures. The main structural components labeled are E: the ellipsoidal shell, R: the ring, E1 and E2: the ears, and six bumps. The structures are described in detail in Section~3.}
\label{OmegaCAM}
\end{figure}

An archival narrow-band image of NGC 6563 was retrieved from the ESO Science Archive (Fig.~\ref{OmegaCAM}). 
The observation was obtained on July 2, 2017, with the VLT Survey Telescope (VST; 2.6-m) at Cerro Paranal, Chile, using the OmegaCAM imager. The dataset corresponds to the ESO programme 177.D-3023(I) and was acquired in imaging mode with the NB-659 filter, whose central wavelength of $\sim \lambda=6590$\,\AA\ and FWHM of $\sim \lambda=100$\,\AA\ samples the {\ha}+{\nii} emission region. The detector was operated in $1\times1$ binning mode, providing a mosaic format of $17152\times16800$ pixels over the full OmegaCAM field of view. The individual exposure time was 120\,s. According to the header, the observation was obtained at an airmass of $\simeq1.09$ and under seeing conditions of about 0.75 arcsec. The image used in this work is the archived science product distributed by ESO. The main morphological features are labelled in this figure and described in Section \ref{sec:EPS}.

In addition to the imaging and spectroscopic datasets described above, we retrieved the astrometric and photometric parameters of NGC\,6563 from the Gaia DR3 archive\footnote{Gaia DR3 archive: https://gea.esac.esa.int/archive/}. The central star (CS) was identified within 1 arcsec as Gaia DR3 source ID 4039600536544395392. However, the CS is not clearly distinguishable against the strong nebular background near the geometric center; thus, its GAIA position is marked with a cross in the RGB image of Figure~1. The Gaia $G$-band magnitude of the CS is $17.276 \pm 0.002$ mag. Previous analyses based on Gaia EDR3 data estimated an effective temperature of 123\,kK, a luminosity of 69.18\,L$_{\odot}$, and a progenitor mass of 2.93\,M$_{\odot}$ \citep{santamaria21}. Additional studies classify NGC\,6563 as a Peimbert Type II PN with near-solar metallicity \citep{Henry}. Using Gaia DR3 photometry and parallax, the CS has also been assigned a 99.99\% probability of being a white dwarf.

\subsection{High-dispersion spectroscopy}
Spectroscopic observations were carried out on 31 May 2025 with the Manchester Echelle Spectrograph (MES; \citealt{meaburn}) mounted on the 2.1-m ($f/7.5$) Arcadio Poveda Telescope at the Observatorio Astron\'omico Nacional in San Pedro M\'artir, Mexico (OAN--SPM). The {\ha}+{\nii} interference filter was employed. A $2\times2$ CCD binning mode was adopted with an E2V-4240 detector ($2048\times2048$ pixels; $13.5\,\mu$m\,pix$^{-1}$), providing a plate scale of 0.35 arcsec\,pix$^{-1}$. The slit length was 6.3\,arcmin and its width was 2\,arcsec (150\,$\mu$m). A $\Delta\lambda=90$\,\AA\ filter isolated the 87$^{\mathrm{th}}$ order, covering the spectral region containing \ha\ and \nii$\lambda6583$. 

\begin{figure}[H]
    \includegraphics[width=0.9\textwidth]{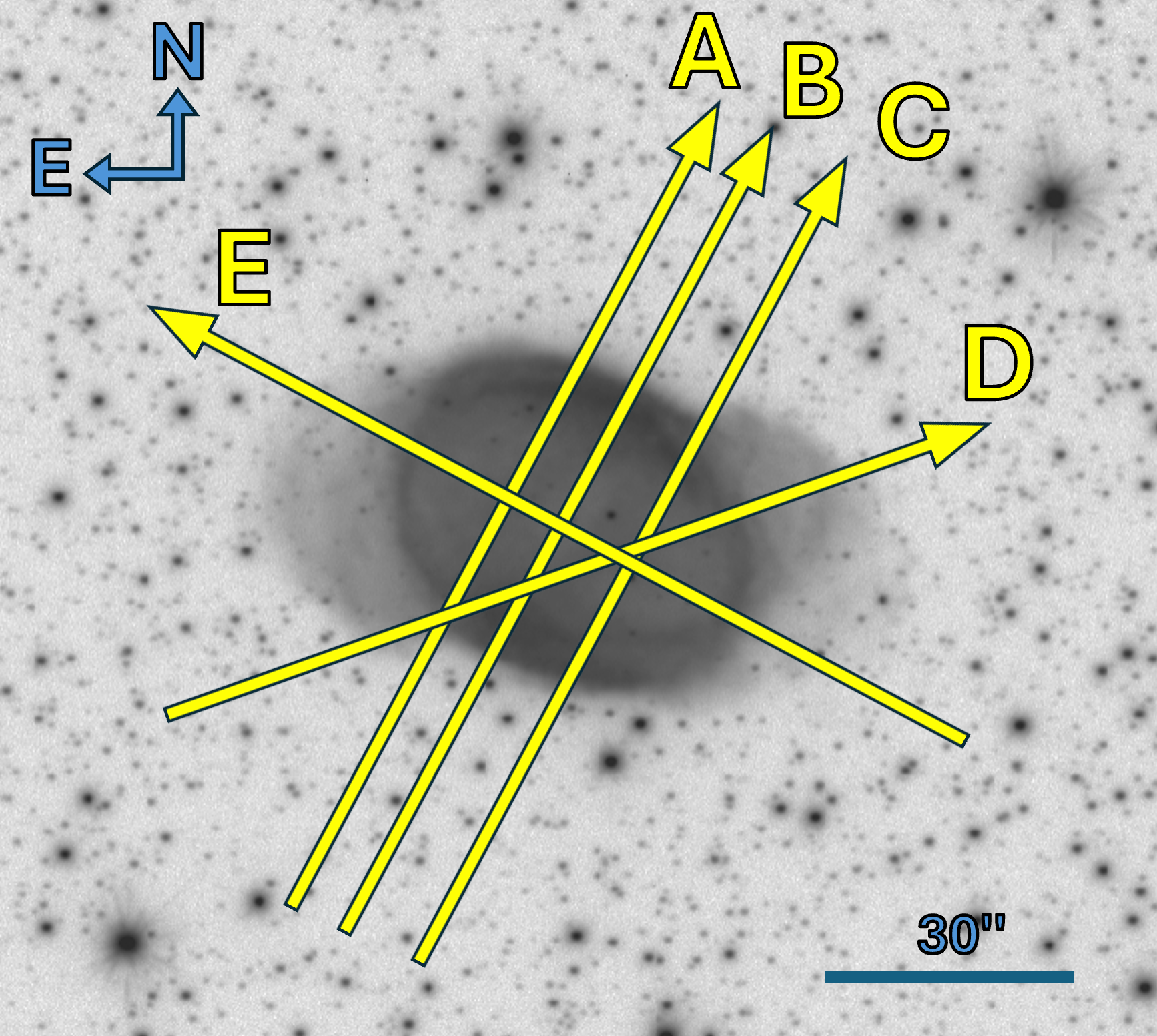}
    \caption{Optical image of NGC~6563 with the slit positions (A--E) overlaid, indicating the nebular regions sampled by the spectroscopic observations. The background image is the same as that shown in Fig.\ref{OmegaCAM}.}
    \label{slits}
\end{figure}

Five slit positions were observed (Fig.~\ref{slits}). The position angles (PAs) for A, B, and C were: $-28$\degr; D: $-70$\degr, and E: +62\degr. The exposure time for slits A, B, and C was 1200\,s, whereas for slits D and E was 1800\,s.

Data reduction was performed following standard procedures for long-slit, high-dispersion spectroscopy using 
the NOIRLab 2.18 version of IRAF, which provides processing routines for long-slit spectroscopy \citep{tody1986,tody1993,fitzpatrick2025}. Wavelength calibration was obtained from ThAr arc-lamp exposures, yielding a spectral scale of 0.057~\AA\,pix$^{-1}$ and an accuracy of approximately $\pm 1$~km~s$^{-1}$. The instrumental spectral resolution, measured as the full width at half maximum (FWHM) of the arc-lamp emission lines, was $\simeq 13$~km~s$^{-1}$. 

\section{Results}
\label{sec:EPS}

\subsection{Morphology}

Using the {\ha}+{\nii} image shown in (Figure~\ref{OmegaCAM}), we identify the principal morphological features of NGC\,6563. The nebula exhibits a well-defined elliptical, egg-like shape (E), with a $\simeq50$\,arcsec major axis at PA$=+65$\degr\, and a $\simeq38$\,arcsec minor axis at PA$=-25$\degr. The two visible “ear-like” extensions (E1 and E2) appear to create an $S$-shaped morphology and are consistent with previous studies \cite{Akashi2021}; however, their boundaries are not clearly resolved in the available data. We therefore assume the two ears to be identical in our analysis. In addition, several small-scale bumps are visible at different locations along the shell. In the northern and southern regions of the ellipsoidal shell, the nebula also shows bright rim-like features, which give the appearance of locally doubled or enhanced shell segments. The possible physical origins of these structures are discussed in Section \ref{section4}.

\subsection{Kinematics}

\begin{figure*}
    \includegraphics[width=\textwidth]{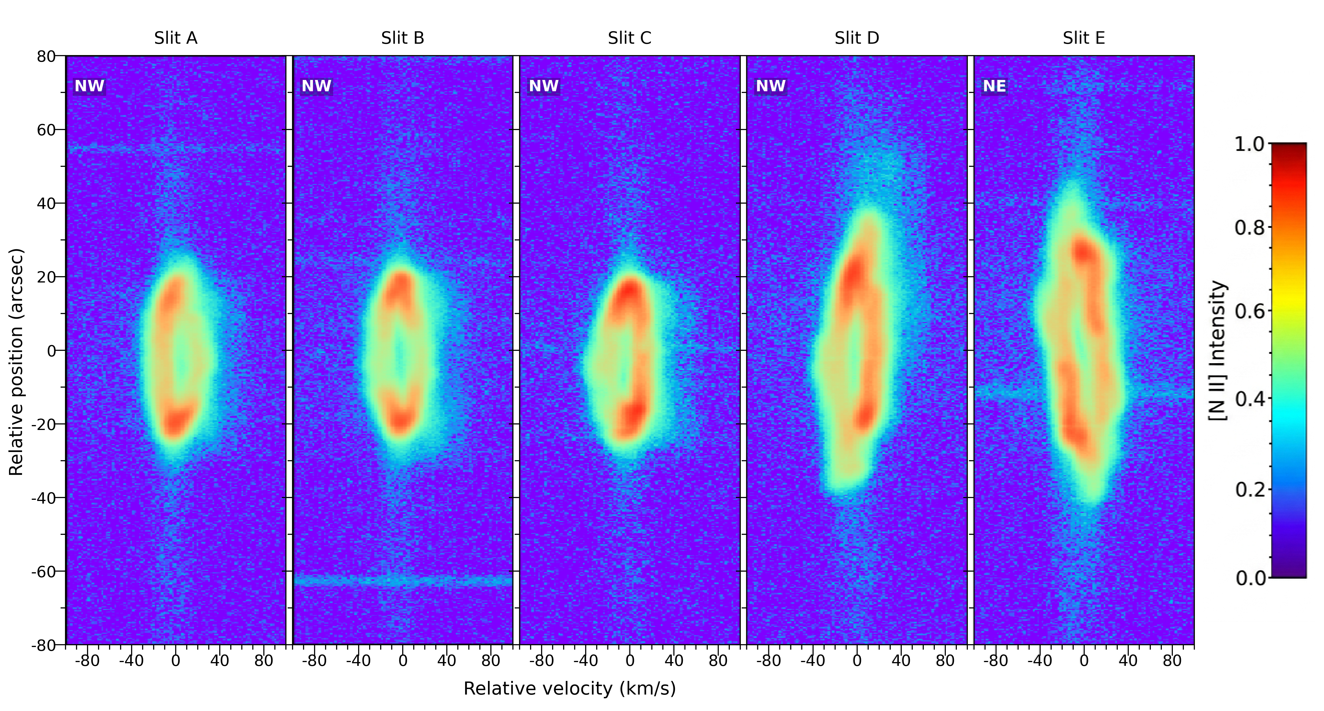}
    \caption{PV diagrams derived from the \nii $\lambda$6583 $\textup{\r{A}}$ emission line. The observed PV diagrams are labelled as A, B, C, D, and E. The color bar represents the normalized \nii $\lambda 6583$ \AA\ intensity.}
    \label{fig:PV}
\end{figure*}

Position–velocity (PV) diagrams were constructed using \texttt{MEZTools}\footnote{https://github.com/mgomezAstro/MEZTools}, a software package that converts pixel coordinates and wavelengths into spatial position and radial velocity, respectively, while taking into account the detector plate scale. The resulting PV diagrams are presented in Figure~\ref{fig:PV} and are described in the following paragraphs. 

$Slit A$: The PV diagram shows a strongly elongated velocity ellipse, typical of a low-velocity ellipsoidal shell. The emission is brighter at the northern and southern extremes, while the central region is fainter. The ellipse appears slightly asymmetric, with the blue side being more intense. E1 is lightly detected at the southern extreme as an extension reaching down to approximately $-20$.
The maximum line splitting is approximately 40\kms. 

$Slit B$: As in Slit A, this PV exhibits a well-defined and elongated velocity ellipse with a comparable velocity extent. The structure is slightly more symmetric than in Slit A. The emission is faintest between $-9$ and $+9$ arcsec. 
The line splitting along this slit is similar to Slit A, reaching approximately 40\kms\ at the central position. 

$Slit C$: The PV diagram of Slit C also shows an elongated velocity ellipse, although it is slightly narrower than those observed in Slits A and B. The line splitting at the central position is approximately 30\kms. The emission is more intense on the red side, while the blue side is significantly fainter between approximately $-20$ and $+5$ arcsec. The emission associated with E2 is visible as a faint northern extension in the observed PV, becoming progressively weaker with increasing distance from the shell. 

$Slit D$: In Slit D, the PV diagram shows a clear point-symmetry, with intense emission concentrated on the red northern side and the blue southern side, indicating a transition from the main shell to the outer, lower-density regions associated with the ears. Both ears are detected, extending to approximately $40$ arcsec for E1 and 
$-40$ arcsec for E2, with fainter emission observed beyond the apparent nebular boundary. E2 exhibits the larger spatial extent. The orientation of both ears remains consistent with their overall $S$-shaped morphology. The line splitting along this slit reaches a maximum of approximately 35\kms\ at a spatial offset of about $3$ arcsec.

$Slit E$: The PV diagram of Slit E also appears like point-symmetric, as the case of Slit D. The emission is concentrated over a broad region on the blue side, between approximately $5$ and $30$ arcsec, and on the red side between about $-5$ and $-30$ arcsec. The width of the velocity ellipse varies along the slit in an asymmetric manner, differing from the behavior observed on the other slits. E1 reaches its largest spatial extent at approximately $42$ arcsec, while E2 extends to about $-42$ arcsec. The central line splitting reaches a value in this slit, of approximately 40\kms. 

The systemic velocity of \ngc\ was determined from the emission-line splitting measured at the geometrical center of the nebula in slits B and E. The midpoint between the two split components was taken as the observed systemic velocity, and the corresponding Heliocentric and Local Standard of Rest (LSR) corrections were then applied. The resulting values for these two slits were averaged as
$V_{\mathrm{sys}}^{\mathrm{LSR}} = -25 \pm 1${\kms} and $V_{\mathrm{sys}}^{\mathrm{Hel}} = -34 \pm 1${\kms}.

\begin{table*}[h!]
\centering
\caption{Distance, angular radius, and kinematic properties of NGC 6563. References in literature are listed in the last column.}
\label{tab:dist}
\begin{tabular}{ccccc}
\hline
Distance & Radius & $V_{\rm sys}^{\rm HEL}$ & $V_{\rm exp}$ & Reference \\
(kpc)    & (arcsec) & (km\,s$^{-1}$) & (km\,s$^{-1}$) & \\
\hline
 $0.93\pm0.11$ & 25 & $-34\pm1$ & $22\pm1$ & This study \\
 $1.84\pm0.74$ & -- & -- & -- & \citep{Hernandez24} \\
 $0.94\pm0.11$ & 29.5 & -- & 21.5 & \citep{santander22} \\
1.01 & 29.5 & -- & -- & \citep{chornay} \\
 $1.67\pm0.33$ & 21.5 & -- & -- & \citep{stangel10} \\
 -- & 29.6 & -- & -- & \citep{tylenda2003} \\
 1.90 & 22.6 & $-21.6$ & -- & \citep{quireza} \\
 -- & 23.8 & -- & -- & \citep{stasi} \\
 1.90 & -- & -- & -- & \citep{maciel}\\
 -- & -- & $-30$ & -- & \citep{beaulieu} \\
 -- & -- & $-29.5\pm3.6$ & -- & \citep{durand} \\
 2.90 & 22.6 & -- & -- & \citep{zhang95} \\
\hline
\end{tabular}
\end{table*}

\vspace{-6pt}

\section{The nature of NGC 6563}\label{section4}

\subsection{Morphokinematic structure}\label{mk}

\begin{figure*}
    \centering
    \includegraphics[width=\textwidth]{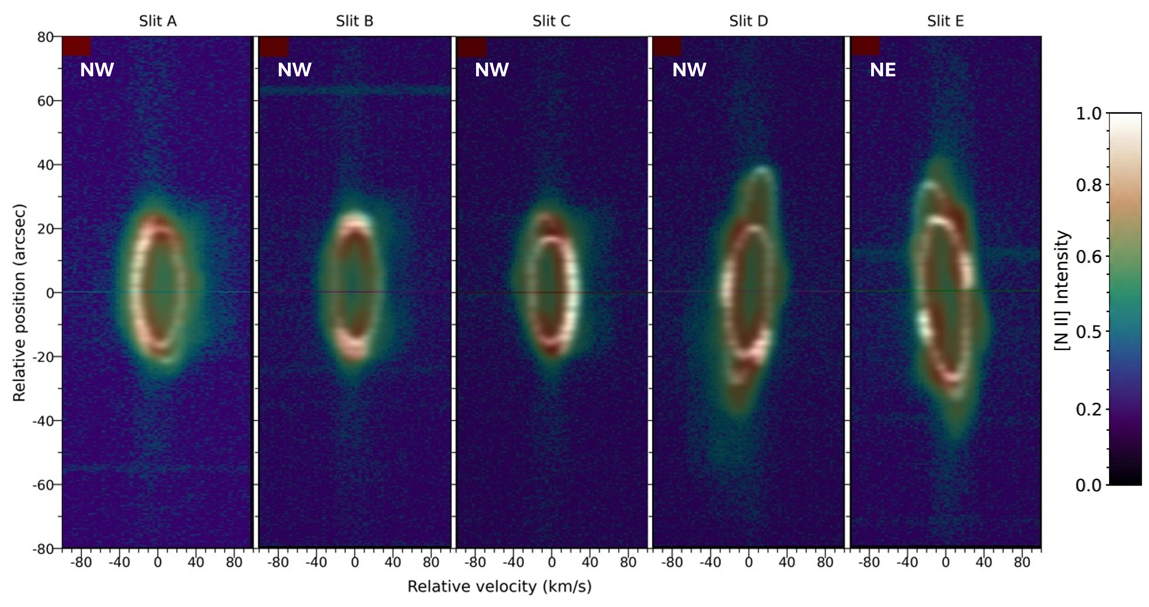}
    \caption{Same as the previous figure, but with the synthetic PV maps from the best-fitting {\sc ShapeX} morpho-kinematic model overplotted on the observed [N II] $\lambda6583$ PV maps for slits A–E across NGC 6563. The model reproduces the main velocity splitting and the overall spatial extent of the nebular emission along the different slit positions. The color bar represents the normalized \nii $\lambda 6583$ \AA\ intensity.}    \label{PVs_model}
\end{figure*}

To determine the structure of {\ngc} more accurately, we constructed a three-dimensional model using \textsc{ShapeX}, a morphokinematic modelling tool for gaseous nebulae \citep{steffen}. The model is constrained using both the PV diagrams and the main images of {\ngc} (Figs.~\ref{rgb}, \ref{OmegaCAM}, and \ref{slits}). The methodology we used to build a model with \textsc{ShapeX} is described in \cite{Vazquez2026} (Appendix B). The resulting PV maps obtained from the best-fit model are shown in Fig.~\ref{PVs_model}. 

Our model comprises four main structural components: a thick ellipsoidal shell with an angular thickness of 2\,arcsec, an equatorial ellipsoidal ring with a radius of 19\,arcsec and a height of 8\,arcsec, and two ear-like protrusions.
The main ellipsoidal shell is characterized by a position angle of 40\degr\ and an inclination of 10\degr. The axis of the ring corresponds to that of the ellipsoidal shell, and its presence helps reproduce the enhanced emission observed in all the PV diagrams remarkably well.
The protrusions, modeled as partial ellipsoidal structures, are hereafter designated E1 and E2, corresponding to the eastern and western extensions, respectively. E1 is oriented at an angle of $45^{\circ}$ relative to the major axis of the main shell, whereas E2 shows a similar inclination on the opposite side.

On the other hand, the [N~II]/H$\alpha$ ratio map shown in Fig.~1 provides an additional morphological diagnostic of NGC~6563. Since this ratio reduces the dominance of structures produced mainly by surface-brightness enhancements, it helps to trace spatial variations in excitation and/or abundance across the nebula. In particular, the ratio map reveals a relatively well-defined, elongated 
high-ratio structure along the outer nebular regions, while the central zone 
shows lower values. This contrast makes the apparent ring-like component more 
evident than in the individual emission-line images. Although this diagnostic 
does not by itself determine the excitation mechanism, it supports the 
identification of the ring-like structure as one of the main morphological 
components considered in our spatio-kinematic model.

\begin{figure*}
	\centering
	\includegraphics[width=\textwidth]{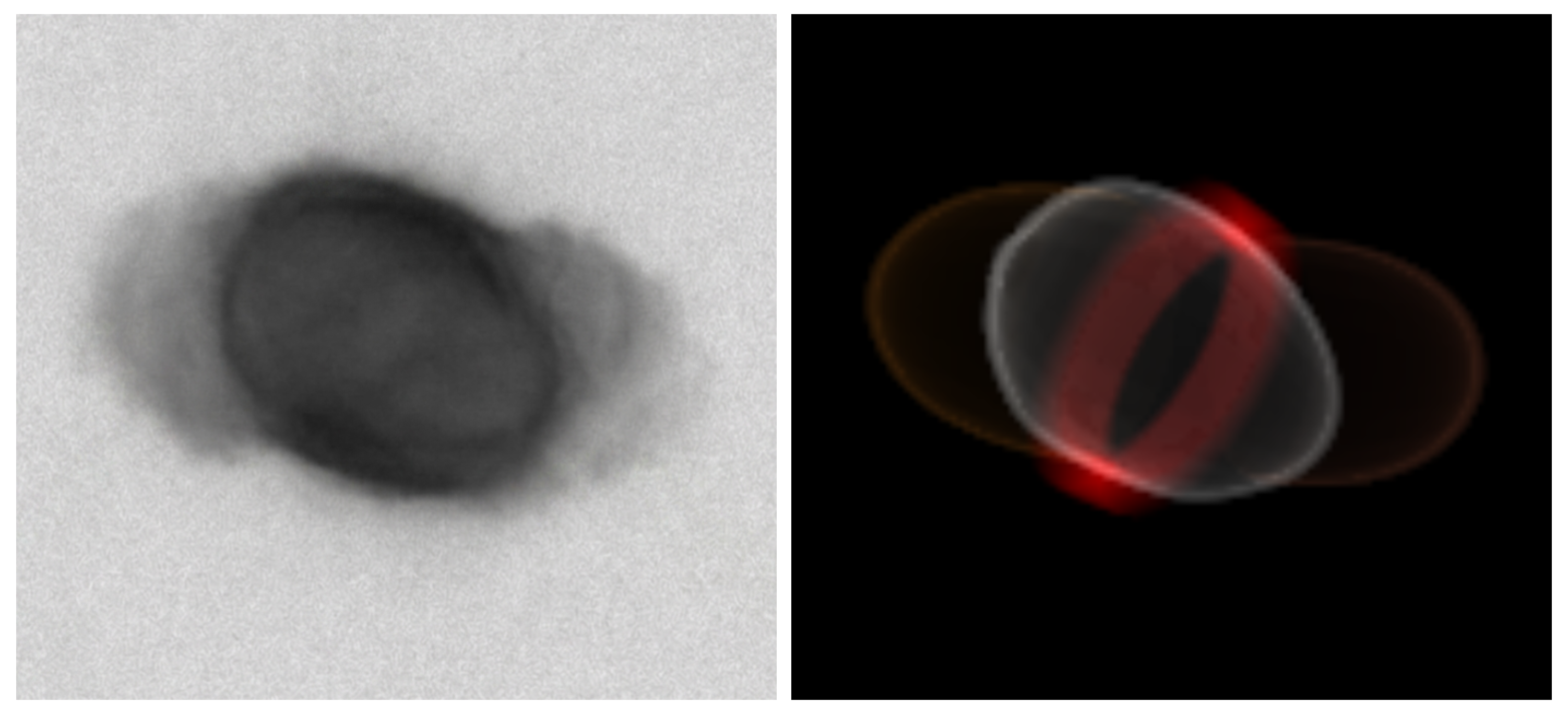}
	\caption{Comparison between the observed and synthetic surface-brightness distributions of NGC 6563. Left: VST/OmegaCAM NB-659 image of NGC 6563. Right: projected surface-brightness distribution generated from the best-fitting SHAPEX morphokinematic model. Both panels are shown with the same angular scale and sky orientation to allow direct comparison between the observed nebular morphology and the model.}    \label{comparison}
\end{figure*}

Furthermore, NGC 6563 exhibits multiple bulge-like structures distributed along the ellipsoidal shell, hereafter referred to as bumps. In order to reproduce the observed morphology, six bumps have been incorporated into the 3D model, as evidenced in the direct images of the main structure (Figs.~\ref{rgb} and \ref{OmegaCAM}). These protrusions are necessary to account for localized intensity enhancements and deviations from a purely smooth ellipsoidal geometry. 
Although the protrusions identified as bumps introduce departures from axial symmetry, they do not define distinct pairs of lobes or independent symmetry axes that would support a multipolar interpretation. Instead, these bumps are better understood as localized shell deformations or density enhancements superposed on the main ellipsoidal structure. This interpretation is supported by Fig.~\ref{comparison}, which presents a direct side-by-side comparison between the observed VST/OmegaCAM image and the synthetic surface-brightness distribution generated from our best-fitting model; the agreement shows that the inclusion of these bumps is sufficient to reproduce the observed nebular morphology without invoking a multipolar structure. 
 
In addition, as part of our morphokinematic interpretation, the bright rim-like features observed toward the northern and southern parts of the shell do not require a second independent ellipsoidal shell. Instead, they can be mainly explained as projected emission from the bases of the ear-like protrusions, possibly combined with local distortions of the outer ellipsoidal shell.

To estimate the kinematic ages of the structures, we followed the approach of \citet{guillen13}, using Eq. \eqref{Eq3}:

\begin{equation}
\tau_{\rm k}=\frac{4\,744r_{\rm p}D}{v_{\rm p}}
\label{Eq3}
\end{equation}

\begin{figure*}
    \centering
    \includegraphics[width=\textwidth]{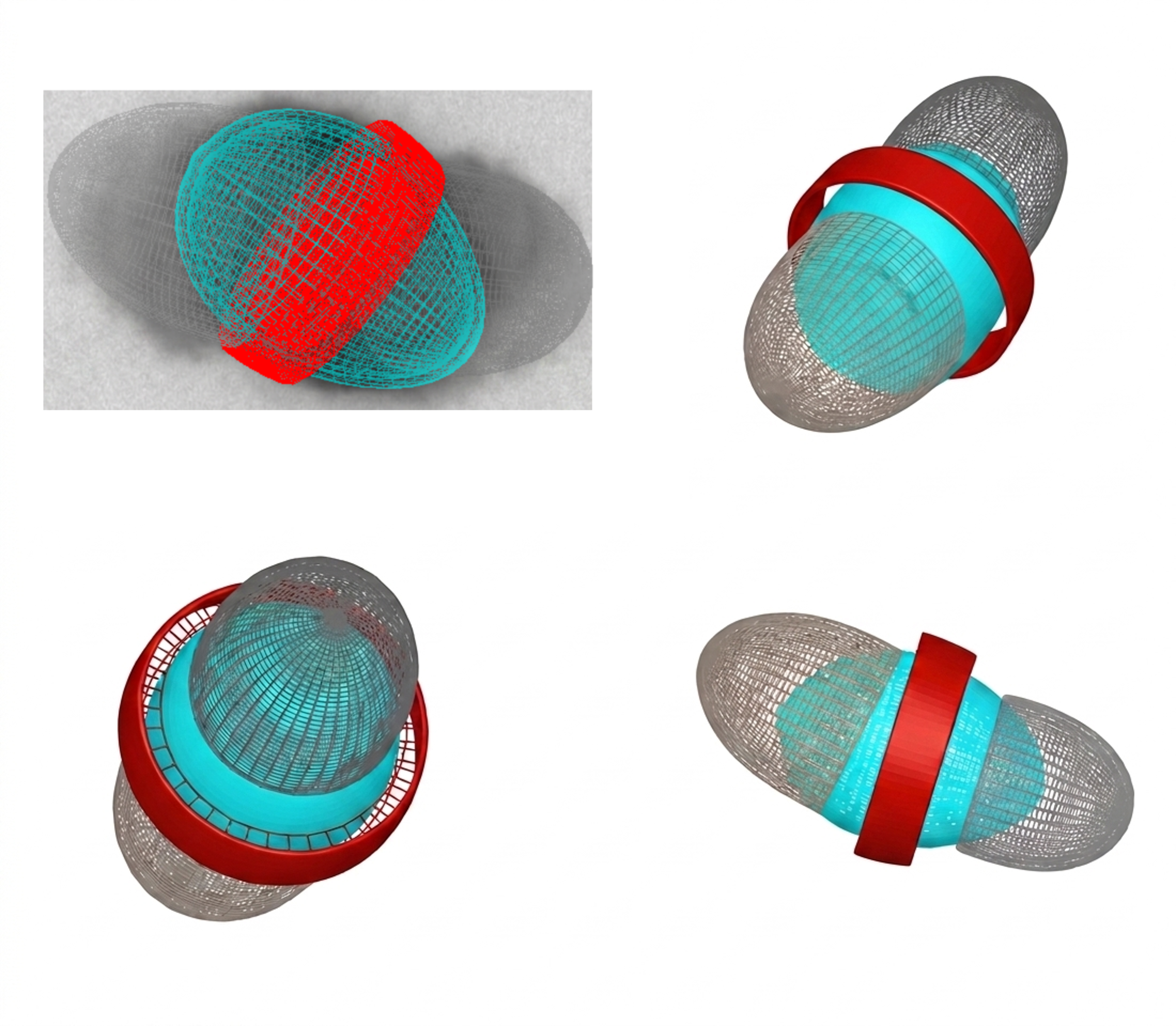}
    \caption{\textsc{ShapeX} morpho-kinematic reconstruction of NGC~6563 constrained by all available PV diagrams. The upper-left panel shows the model from the observer’s viewpoint, overlaid on the original image to allow a direct comparison with the observed morphology, while the remaining panels present the same reconstruction arbitrarily rotated about the $x$-, $y$-, and $z$-axes, respectively, to highlight its three-dimensional structure. The cyan structure represents the ellipsoidal shell with six integrated bumps, the upper and lower ears are represented by the brown/gray semi-ellipsoidal structures, and the red belt represents the ring.
    }    \label{shapexx}
\end{figure*}

\noindent Here, $r_{\rm p}$ (arcsec) represents the polar radius, $v_{\rm p}$ (\kms) the polar expansion velocity, and $D$ (kpc) the distance to the PN. The factor 4744 results from unit conversion and ensures consistency between angular measurements, physical distances, velocities, and time when expressed in the International System of Units (MKS). 
To determine the size and velocities of each structure we used the model shown in Fig.~\ref{shapexx}, moving the PAs and inclination angles to set each structure in a convenient position and take synthetic PVs to make measurements.

In this study, we adopt a distance based on the Gaia DR3 trigonometric parallax of $\varpi = 1.07\pm0.13$ mas \citep{gaiadr3}, yielding $d = 0.93\pm0.11$ kpc (see Table~\ref{tab:params}). \cite{santander22} and \cite{chornay} also used Gaia trigonometric parallaxes to determine distances to PNe. Table~\ref{tab:dist} shows that previous studies reported larger distances for {\ngc}. For example, \cite{zhang95}, \cite{Stan2010}, and \cite{durand} calibrated distances using PNe with known masses, temperatures, and distances, while \cite{maciel} relied on multiple catalogues to derive distance estimates. \cite{quireza} combined parallaxes, expansion methods, and literature values, and \cite{Hernandez24} employed both Gaia parallaxes and statistical distance scales. Given that Gaia provides a relatively precise parallax for {\ngc}, with an uncertainty of 0.13 mas, we consider adopting a parallax-based distance using $d = 1/\varpi$ to be a reliable and appropriate choice for this study.

We derive kinematic ages of $\tau_{\mathrm{k}} = 3\,700 \pm 700$ yr for both the ellipsoid and the ring, $\tau_{\mathrm{k}} = 7\,500 \pm 1\,000$ yr for E1, and $\tau_{\mathrm{k}} = 8\,800 \pm 1\,500$ yr for E2. The identified structures, sizes, velocities, and corresponding kinematic ages are listed in Table~\ref{tab:kin}. These values indicate that both ears are older than the main nebular shell \citep{akashi}. Since $\tau_{\mathrm{k}}$ scales linearly with distance, adopting larger distances from the literature would proportionally increase the inferred ages. For example, for distances of 1.67, 1.84, and 2.90 kpc, the kinematic age of the main shell would increase by 80\%, 98\%, and 212\%, respectively, reaching values from about 6\,600 yr to 11\,500 yr.

In evaluating our current data, we must note that the available imaging spans a baseline of only ~8 years (2017–2025). At a distance of 0.93 kpc and expansion velocities of 20–30 km s$^{-1}$, the expected angular expansion is only a few hundredths of an arcsecond. This displacement falls below our robust detection threshold, particularly when accounting for discrepancies between different instruments, filters, point spread functions (PSFs), and varying spatial resolutions. Consequently, while our $\tau_{\mathrm{k}} = r/v $ estimates provide a reasonable kinematic framework, a direct measurement of expansion remains beyond the reach of the current multi-epoch dataset.

In the next subsection, we discuss the possible physical mechanisms responsible for the observed morphologies and kinematic ages of these structures.

\begin{table}
\centering
\caption{The labels and structures in the first and second columns correspond to the components modelled with \textsc{ShapeX} (Fig.~\ref{shapexx}). The quantities $v$ and $r$ represent the expansion velocity in km\,s$^{-1}$ and the angular distance from the nebular centre in arcsec, respectively, and may refer to either the polar or equatorial direction depending on the structure. The parameter $k$ is the proportionality constant of the homologous expansion law adopted for each structure, $v = kr$. The parameter $\tau_{\mathrm{k}}$ denotes the kinematic age, rounded to the nearest hundred years.}
\renewcommand{\arraystretch}{1.5}
\begin{tabular}{lllcccc}
 \hline \hline 
Label & Structure & Orientation & $v$    & $r$      & $k$                         & $\tau_{\rm k}$ \\ 
      &           &             & [\kms] & [arcsec] & [km s$^{-1}$ arcsec$^{-1}$] & [yr] \\ 
\hline
E & Ellipsoid        & polar &  $30\pm1$ & $25\pm1$  &  1.2 &  $3\,700\pm700$   \\ 
   &            & equatorial &  $22\pm1$ & $18\pm1$  &  1.2 &  $3\,700\pm700$   \\ 
R  & Ring       & equatorial &  $23\pm1$ & $19\pm1$  &  1.2 &  $3\,700\pm700$   \\ 
E1   & East ear      & polar &  $20\pm1$ & $34\pm1$  &  1 &  $7\,500\pm1\,000$  \\ 
     &          & equatorial & $10\pm1$ & $17\pm1$  &   1 &  $7\,500\pm1\,000$  \\ 
E2   & West ear      & polar &  $20\pm1$ & $40\pm1$  &   1      &  $8\,800\pm1\,500$   \\ 
    &          & equatorial  &  $10\pm1$ & $20\pm1$   &   1      &  $8\,800\pm1\,500$   \\ \hline

\end{tabular}  
\label{tab:kin}
\end{table}

\subsection{Origin and Shaping of the Nebular Morphology}

The elliptical and slightly egg-shaped morphology of NGC 6563 can be interpreted within established shaping mechanisms for PNe. Early hydrodynamical models reproduced elliptical structures using {\it ad hoc} prescriptions for the asymptotic giant branch (AGB) wind density distribution \citep{icke89,mellema91,frank94}, while later studies incorporated stellar rotation \citep{garcia99}. Magnetized interacting-wind models further showed that even a spherically symmetric slow AGB wind can evolve into an elliptical or weakly bipolar nebula when the fast post-AGB wind carries a magnetic field \citep{Chevalier,rozyczka96}. In these models, magnetic tension in the shocked wind region redirects material toward the polar directions. An anticorrelation between progenitor mass and magnetic collimation efficiency has been suggested, with elliptical nebulae typically associated with lower-mass progenitors where magnetic shaping dominates \citep{garcia99}.

Binary interaction may also contribute to the observed asymmetry \citep{ondra}. \cite{soker89} proposed that a close companion emerging from a common-envelope phase can eject material preferentially in the equatorial plane, producing density enhancements and localised structures. Such interactions can generate asymmetric inner regions and knot-like features. Moreover, tidal interaction between an AGB star and a low-mass companion may enhance inner asymmetries, a common characteristic of elliptical PNe \citep{soker95}. Therefore, the morphology of NGC 6563 is most consistently interpreted within a jet-driven shaping scenario linked to binary interaction. In this framework, any magnetic fields contributing to collimation would be associated with the companion’s accretion disk and/or the launched collimated outflows, rather than with a stationary field from a single rotating AGB star. This scenario naturally accounts for the point-symmetric structure and the inferred change in the outflow axis.

The derived kinematic ages of the different structures in \ngc\ are consistent with the evolutionary scenario proposed by previous studies (e.g. \cite{schwarz92}, \cite{tocknell2014}, \cite{Akashi2021}). The upper (E1) and lower (E2) ears have kinematic ages ranging from $7\,500\pm1000$ to $8\,800\pm1500$ yr, significantly older than the main ellipsoid and shell, which have an age of $3700\pm700$ yr. This age difference supports the idea that the ears were formed prior to the development of the main nebular shell. We note, however, that the kinematic ages of the ears assume ballistic expansion at constant velocity. If the ear-like structures have been significantly decelerated by interaction with the post-AGB wind, previously ejected nebular material, or the ambient interstellar medium, their $\tau_{\rm k} = r/v$ ages should be regarded as upper limits.

Both ears appear largely symmetric and morphologically similar, in agreement with previous studies \citep{schwarz92, Akashi2021}. Another way to interpret the age differences in their spatial extent, kinematic ages, and expansion velocities, may be the deviations from homologous expansion and/or asymmetric interactions within the system. The apparent precession of the ears (see Fig.~\ref{shapexx}) further supports a binary interaction scenario, which can naturally account for the observed morphology and kinematic asymmetries \citep{ondra, lopez22}.

According to the scenario proposed by \cite{Akashi2021}, short-lived jets launched by a companion star interact with the regular AGB wind, shaping the ears. These jets are produced when the companion accretes mass from the AGB progenitor and subsequently ejects collimated outflows. At a later stage, the system undergoes common-envelope (CE) evolution, which leads to the ejection of a dense equatorial shell that forms the main body of the PN. In this framework, the jets—and consequently the ears—precede the formation of the dense main shell and interact primarily with the earlier, lower-density AGB wind. In this context, hydrodynamical simulations have shown that bipolar jets can also compress circumstellar material toward the equatorial plane, naturally leading to the formation of a dense equatorial ring (e.g., \citealt{akashi15}). This provides a unified framework in which the ears and the ring are both linked to binary interaction and jet activity, although they were likely formed during different evolutionary stages. Complementary simulations by \cite{lopez22} have shown that jets may also be launched during the CE phase itself, when the companion is embedded within the envelope. Depending on the jet power and the local envelope conditions, such jets may be choked or only partially break out, naturally leading to modest bipolar protrusions rather than extended lobes. Both scenarios highlight the role of binary interaction and jet activity in producing elliptical nebulae with ear-like structures, and may therefore provide a plausible framework for interpreting the morphology of NGC 6563. While \cite{soker22} proposed that 'ears' in supernova remnants and similar nebulae may result from late jet-launching episodes that penetrate the main shell, this model is not in contradiction with the present study. Rather, these diverse scenarios highlight the wide variety of environments in which jets can interact with previously ejected circumstellar material. In the case of NGC 6563, our kinematic analysis reveals that the ears possess a larger kinematic age than the main nebula, indicating they were formed by early jet activity—likely during a pre-common envelope phase—rather than by late-stage pulses. Thus, while late jet–CSM interactions provide a viable mechanism for producing ear-like protrusions in other astrophysical environments, the kinematic ordering inferred for NGC~6563 favours an early jet–AGB-wind interaction, as proposed for ear formation in PNe by \cite{Akashi2021}.

The PV diagrams reveal a kinematic asymmetry, with the right side exhibiting larger expansion velocities compared to the opposite side. Such velocity imbalances are commonly observed in evolved PNe and may reflect density gradients in the surrounding medium or intrinsic asymmetries in the mass-loss history of the progenitor star  \citep{kahn1983, dwark}. A faster expansion on one side could indicate reduced external pressure in that direction, allowing the shell to expand more freely.

The enhanced emission intensity regions seen in the PV diagrams likely correspond to density enhancements within the nebular shell or zones of compression. Although the ring is closely aligned with the waist of the ellipsoidal structure in projection, the ShapeX morphokinematic modelling indicates that a separate ring component is required to reproduce the observed PV signatures and localized intensity enhancements. Such features may arise from hydrodynamic instabilities in thin expanding shells \citep{vishniac} or from the interaction between the fast wind and previously ejected material \citep{dwark}. In addition, our 3D morpho-kinematic modeling demonstrates that a significant fraction of the observed enhancement originates from the equatorial ring, whose higher density produces a pronounced signature in the PV diagrams.

The observed kinematic asymmetry may indicate interaction with a non-uniform ambient medium. If the western rim is expanding more slowly, this could suggest enhanced external pressure in that direction, consistent with interaction with the surrounding interstellar medium. Similar behaviour has been reported in evolved PNe exhibiting ISM interaction signatures \citep{wareing}.

The irregular surface of NGC 6563, characterised by several bumps distributed along the ellipsoidal shell, suggests that the nebular envelope is not dynamically smooth. These deviations from a regular geometry likely trace localized compressions within the shell, indicating that the gas density is structured rather than homogeneous. Such surface distortions are frequently observed in evolved PNe and can develop naturally in dynamically evolving, pressure-confined shells.

If the expanding envelope of NGC 6563 experiences significant compression, either from internal wind interactions or external confinement, it may become dynamically unstable. In particular, thin radiative shells are prone to growth of perturbations when velocity shear or pressure gradients are present. Under these conditions, small-scale irregularities can amplify over time, giving rise to bubble-like protrusions and localized thickening of the shell. This behaviour is consistent with instability mechanisms operating in shocked layers, including the non-linear thin-shell instability described by \cite{2025MNRAS.541.3932F,vishniac}.

Within the context of the interacting stellar winds scenario \citep{kahn,dwark}, the fast post-AGB wind sweeping up the earlier slow AGB outflow generates a compressed shell bounded by shocks. The resulting velocity gradients across this interface can enhance departures from spherical symmetry, particularly if the density of the swept-up material varies azimuthally. In this framework, the observed bumps in NGC 6563 can be interpreted as dynamical responses of a compressed shell to internal wind interaction rather than signatures of discrete episodic ejections.

Moreover, interaction with a non-uniform interstellar medium may further contribute to the development or amplification of these structures. If the nebula expands into an ambient medium with density gradients, differential deceleration across the shell can introduce additional distortions and localized enhancements. Such combined effects — internal wind interaction and external ISM influence — provide a plausible explanation for why pronounced protrusions are required in the morphokinematic reconstruction, while appearing comparatively subtle in direct imaging due to projection effects and line-of-sight integration.
\section{Conclusions}
\noindent

Within this work, we analysed the morphology and kinematics of \ngc\ using high-resolution imaging and spectroscopy. From the MEZCAL spectra, we derived a systemic velocity of $V_{\rm sys}^{\rm LSR} = -25 \pm 1$\kms\ ($V_{\rm sys}^{\rm Hel} = -34 \pm 1$\kms) and an expansion velocity of the main shell of $V_{\rm exp} = 22 \pm 1$\kms. Combining direct imaging with PV diagrams, we constructed a three-dimensional morphokinematic model consisting of a main ellipsoidal structure surrounded by an enhanced ring, two ear-like protrusions, and six localized bumps along the line of sight. The presence of an enhanced equatorial ring, clearly identified in both the PV diagrams and the 3D morphokinematic reconstruction, indicates a density contrast within the nebula, consistent with compression of material in the equatorial plane. The [N~II]/H$\alpha$ ratio map provides an additional diagnostic of the nebular morphology and excitation structure. This map shows enhanced values along the outer nebular regions, reaching values up to $\sim$1.31. Although the full extent of the ear-like structures is not clearly traced in this map, the enhanced peripheral ratios indicate significant low-ionization emission and may suggest localized excitation variations. However, this diagnostic alone does not allow us to distinguish unambiguously between photoionization and a possible shock contribution.

The reconstructed geometry yields a kinematic age of $3\,700 \pm 700$ yr for the ellipsoidal shell and ring, while the ears are significantly older, ranging from $7\,500 \pm 1\,000$ to $8\,800 \pm 1\,500$ yr. These estimates assume ballistic, homologous expansion at constant velocity. If the material has been significantly decelerated, whether by interaction with nebular gas, the post-AGB wind, or the ambient ISM, the simple $t = r/v$ estimate likely overestimates the true time since ejection. Furthermore, these ages are highly sensitive to the adopted distance, since variations in the distance scale would shift these timescales linearly. Finally, these ages reflect the currently detectable extent of the structures; should future, higher-sensitivity observations reveal more extended emission, the inferred kinematic ages would increase accordingly.

The significantly older age of the ears indicates that they predate the formation of the main nebular shell. This supports a scenario in which the ears originated from earlier collimated outflows, likely associated with a binary interaction phase preceding the ejection of the dense shell. The point-symmetric arrangement of the ears, together with their different spatial extents and velocities, supports a jet-driven scenario in which the outflow axis changed with time.

The observed kinematic asymmetry, with one side expanding faster than the opposite side, together with localised intensity enhancements and surface distortions, suggests that NGC 6563 is evolving within a non-uniform ambient medium. Such differential expansion can arise from asymmetric deceleration caused by density gradients in the surrounding interstellar medium. At the same time, the overall elliptical morphology and the presence of older ear-like protrusions indicate that additional shaping mechanisms have likely been at work during earlier evolutionary phases. Internal wind interaction, thin-shell instabilities, binary-driven mass-loss episodes, and environmental effects may therefore all contribute to the present morphology, operating at different epochs and spatial scales. Further observational and theoretical studies will be necessary to better constrain the relative contributions of these mechanisms.

\vspace{6pt} 

\authorcontributions{Conceptualization, Z.A, R.V.; methodology, Z.A., R.V., F.S.-B.; validation, Y. K., G.R.-L.; formal analysis, Z.A., R.V.; investigation, Z.A., R.V., F.S.-B.; resources, R.V., 
F.S.-B., G.R.-L.; data curation, Z.A., R.V., F.S.-B.; writing---original draft preparation, Z.A., R.V., Y.K.; writing---review and editing, G.R.-L.; visualization, Z.A., F.S.-B.; supervision, Y.K., G.R.-L.; project administration, R.V.; funding acquisition, R.V. All authors have read and agreed to the published version of the manuscript.
}

\funding{Z.A. gratefully acknowledges financial support from İlim Yayma Vakfı (İYV), T\"urkiye. This work was supported by UNAM-PAPIIT IN103125 grant (Mexico). G.R.-L. acknowledges support from SECIHTI, Mexico (grant CBF-2026-843).}

\dataavailability{The data files will be shared on request to the first author.}

\acknowledgments{\textls[-15]
	{This study is based upon observations carried out at the Observatorio Astronómico Nacional in the Sierra San Pedro Mártir (OAN-SPM), Baja California, Mexico. The authors thank the OAN-SPM staff, particularly the telescope operators Gustavo Melgoza (``Tiky''), Francisco Guillén (``Paco Beretta''), Felipe Montalvo, María Riesgo, and Salvador Monroy (``Capt. Storm''). This work is also based on observations collected at the European Southern Observatory under ESO programme 177.D-3023(I). We thank the reviewers for their careful reading and constructive comments, which have helped us to improve the manuscript. The authors also thank Nico Koning for his assistance with the {\sc ShapeX} software. This work has made use of results from the European Space Agency (ESA) mission Gaia, whose data were processed by the Gaia Data Processing and Analysis Consortium (DPAC). Funding for the DPAC has been provided by national institutions, in particular those participating in the Gaia Multilateral Agreement. This research has also made use of the SIMBAD database and the VizieR catalogue access tool, operated at CDS, Strasbourg, France. During the preparation of this manuscript, the authors used ChatGPT 5.5 for the purposes of grammatical revision. The authors have reviewed and edited the output and take full
		responsibility for the content of this publication.}}

\conflictsofinterest{The authors declare no conflicts of interest.} 


\
\begin{adjustwidth}{-\extralength}{0cm}

\reftitle{References}


\bibliography{example.bib}

\PublishersNote{}
\end{adjustwidth}
\end{document}